\newcommand{\flux}{\hbox{erg~cm$^{-2}$~s$^{-1}$}}
\newcommand{\lumin}{{erg~s$^{-1}$}}

\newcommand{\simgt}{\lower 2pt \hbox{$\, \buildrel {\scriptstyle >}\over {\scriptstyle\sim}\,$}}
\newcommand{\simlt}{\lower 2pt \hbox{$\, \buildrel {\scriptstyle <}\over {\scriptstyle\sim}\,$}}

\newcommand{\chandra}{{\emph{Chandra}}}

\documentclass[12pt,preprint]{aastex}

\received{2005 September 6}
\begin{document}


\title{\boldmath$Chandra$ X-ray Observations of Galaxies in an Off-Center Region of the Coma Cluster}


\author{A.E.~Hornschemeier,\altaffilmark{1,2} 
B.~Mobasher\altaffilmark{3} 
D.M.~Alexander,\altaffilmark{4} 
F.E.~Bauer,\altaffilmark{5}\\
M.W.~Bautz,\altaffilmark{6} 
D.~Hammer,\altaffilmark{1} 
B.M.~Poggianti\altaffilmark{7}
}

\altaffiltext{1}{Laboratory for X-ray Astrophysics, NASA Goddard Space Flight Center, Code 662.0, Greenbelt, MD 20771}
\altaffiltext{2}{Johns Hopkins University, 3400 N. Charles Street, Baltimore, MD 21218}
\altaffiltext{3}{Space Telescope Science Institute, 3700 San Martin Drive, Baltimore, MD 21218}
\altaffiltext{4}{Institute of Astronomy, Madingley Road, Cambridge, CB3 0HA, UK}
\altaffiltext{5}{Chandra Felllow, Columbia University, 550 W. 112th Street,
New York, NY 10027}
\altaffiltext{6}{Kavli Institute for Astrophysics and Space Research, 
 Massachusetts
Institute of Technology, 77 Massachusetts Avenue, Cambridge, MA 02139}
\altaffiltext{7}{ Osservatorio Astronomico di Padova, Vicolo dell'Osservatorio 5, 35122 Padua, Italy}

%
%
%
%
%
%


\begin{abstract}

We have performed a pilot $Chandra$ survey of an off-center region of
the Coma cluster to explore the X-ray properties 
and Luminosity Function of normal galaxies.   We present 
results on 13 $Chandra$-detected galaxies with optical photometric matches, 
including four spectroscopically-confirmed Coma-member 
galaxies.   
All seven spectroscopically confirmed giant Coma galaxies in this field have detections
or limits consistent with low X-ray to optical flux ratios (${f_{X}\over{f_{R}}}<10^{-3}$).
We do not have sufficient numbers of X-ray detected galaxies to directly measure
the galaxy X-ray Luminosity Function (XLF).  However, since we have a well-measured
{\it optical} LF, we take this low X-ray to optical flux ratio for the 7 spectroscopically
confirmed galaxies to translate the optical LF to an XLF.  We find good agreement
with Finoguenov et al. (2004), indicating that the X-ray emission per unit optical
flux per galaxy is suppressed in clusters of galaxies, but 
extends this work to a specific off-center environment in the Coma cluster.
Finally, we report the discovery of a region of diffuse X-ray flux which might  
correspond to a small group interacting with the Coma Intra-Cluster Medium (ICM).

\end{abstract}




\setcounter{footnote}{5}

\section{Introduction \label{intro}}

The launch of new X-ray observatories over the last few years has extended the
study of X-ray emission from galaxies due to accreting binaries 
and hot interstellar medium (ISM) to cosmologically interesting distances. 
\citep[e.g.,][]{Horn01}.  Recently, the first X-ray Luminosity Function (XLF) for
normal galaxies (non-AGNs) was constructed at $z\approx0.3$ and $z\approx0.7$
\citep[][]{Norman2004,Ptak2006} using data from the 
two deepest extragalactic 
X-ray surveys \citep[e.g.,][]{davocatalog}. Moreover, a number of major studies
from {\it observationally complete} samples (i.e., observed to some fixed luminosity sensitivity) of 
X-ray detected/selected normal/star-forming galaxies have revealed that 
X-ray emission closely traces star formation rate in galaxies 
\citep[e.g.,][]{BauerXII,Ranalli03,GrimmSFR,Georgakakis04}.

We have yet to explore galaxies in their most typical environments in the
X-ray band -- this is the cluster and group environment, where most of
the galaxies in the current universe are found \citep[][]{Mulchaey03}.  
Also, in order to understand the X-ray properties of galaxies at 
relatively high redshifts, 
one needs a nearby control sample, with a well-known selection function. Until
very recently, this was not possible as there have been relatively few 
complete samples of X-ray selected galaxies assembled at $z\simlt0.1$. 
Recent work using wide-field optical surveys combined with $Chandra$ and 
$XMM$-$Newton$ archival data \citep{Georgakakis04,HornSDSS} have reached 
 $z\approx0.1$ while the $ROSAT$ All-Sky Survey data are expected to 
provide an estimate of statistical properties of X-ray sources and their
Luminosity Function (LF) in the local Universe\citep[$\simlt100$~Mpc; ][]{Tajer05}.  These surveys necessarily cover very large solid angles on the sky (several square degrees and larger) which require many pointings by the relevant
X-ray missions (whose fields of view are typically 0.1-0.2 square degrees).

To understand X-ray properties of local galaxies, we employ
a different strategy than the wider-field X-ray galaxy studies
 \citep{Georgakakis04,
HornSDSS,Tajer05} by performing a pilot $Chandra$ survey in the outskirts of 
the nearest rich cluster of galaxies, the Coma cluster \citep[$z=0.023$;][]
{Colless1996}.   The field is located $\approx41$~arcmin ($\approx1.2$~Mpc,$\approx0.4$ virial radii)
 from the cluster center \citep[see Figure 1; the virial radius is 2.9 Mpc, 
assuming $H_{0}=70$ km s$^{-1}$ Mpc$^{-1}$][]{Lokas2003}.  We target a cluster
due to the large number of galaxies, making the X-ray observations more efficient.
The reason for targeting a region away from the center 
of Coma is the lower X-ray intracluster medium (ICM) surface brightness there.
One of our main goals is to assemble the XLF
for normal galaxies down to the faintest X-ray limits possible.  
Thus far,  X-ray luminosity functions have been assembled in the 
field at high redshift 
\citep[e.g,][]{Norman2004,Ptak2006}, 
for elliptical galaxies in the nearby Coma cluster of galaxies \citep{Fino2004}
and in an XMM-Newton survey of 2dF/SDSS fields at $z\approx0.1$ \citep{Georg2005}. 
$Chandra$ is ideal for this purpose, 
as it allows us to unambiguously resolve the galaxies from any
residual ICM background.  

The $Chandra$ observations cover approximately 20\% of the full 
$\approx 30^{\prime} \times 50^{\prime}$ area in this part of Coma that
has been under intense study. 
This region, and a corresponding region at the center of Coma,
 has extensive optical photometric and medium resolution (6--9\AA)
spectroscopic data for a well-defined sample of giant and dwarf galaxies 
\citep{Mobasher2001spectra,Pogg2001spectra}. As a result, the {\it optical} 
LFs for galaxies at both the core and outskirts of the Coma cluster was
constructed. This allows for a direct comparison between the optical and
X-ray LFs established for the same field. 

Recently, using the XLF for bright ellipticals (over a large 
area) in the Coma cluster it was shown that  
the X-ray emission from Coma galaxies has undergone adiabatic compression 
by the surrounding ICM \citep{Fino2004ICM}.  Furthermore, it was demonstrated
that the X-ray activity of Coma-member galaxies is suppressed with respect 
to the field by a factor of 5.6 \citep{Fino2004}.  However, detailed analyses of these elliptical
galaxies, which were mainly located interior to our field (within 1~Mpc of
the center of Coma), showed that their X-ray emission, although showing
evidence of some compression by the surrounding ICM, was not largely different
from field galaxies \citep{Fino2004ICM}.    \cite{Fino2004} suggested
that it was not so much that the X-ray luminosity was suppressed but that
the X-ray/optical flux {\it ratio} was suppressed.   The reasons for this
might include past gas stripping and/or past suppression of star
formation in these Coma galaxies.

Throughout this paper we assume $H_{0}=70$~km s$^{-1}$ Mpc$^{-1}$. 
The Galactic neutral hydrogen column density is low in the 
direction of Coma 
\citep[$N_{\rm H} = 9.2 \times10^{19}$~cm$^{-2}$; ][]{Stark92}.  
Finally, throughout this paper, we assume a distance modulus of 
35.13 magnitudes \citep{Mobasher2003LF}.

\section{$Chandra$ observation of the Coma cluster of galaxies}
\label{chandraobs}

\begin{figure}[]
\centerline{\includegraphics[scale=0.80,angle=0]{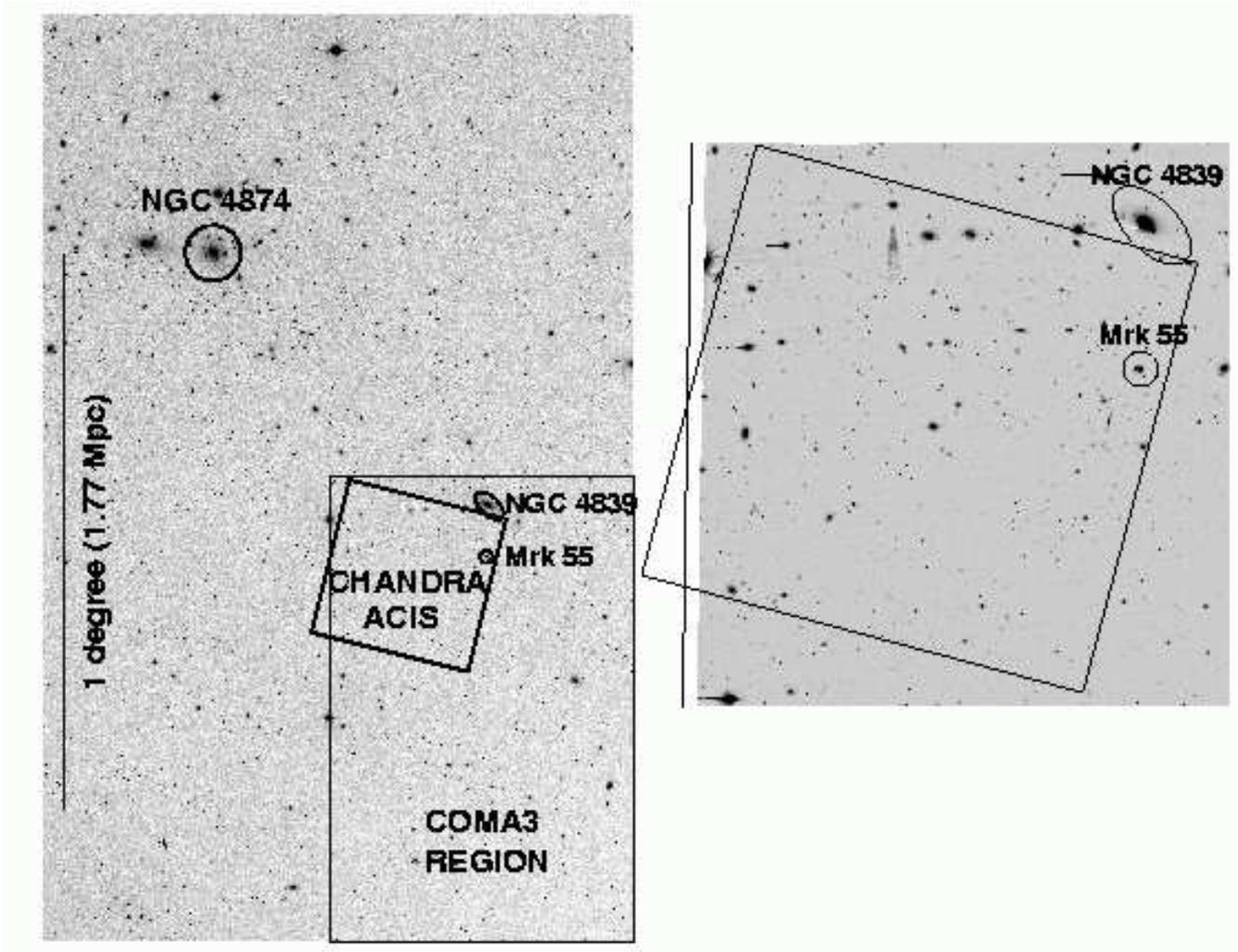}}
\caption[]{ \label{big_image}{(\bf RIGHT:)} {\bf NOTE: This figure is at REDUCED
resolution for the astro-ph version of this paper.}  Large $R$-band image from the
Palomar Digital Sky Survey showing the location of the $Chandra$ ACIS field
within the Coma cluster.  {(\bf LEFT:)} Deeper $R$-band image from the 
William Herschel Telescope showing only the region covered by this survey.
The center of the $Chandra$ ACIS field is 41 arcminutes from NGC 4874, which
is located at the center of the Coma cluster.   For reference, an oval marks the location of
NGC~4839, which is the primary galaxy in a galaxy group falling into the center of 
Coma.  A smaller circle marks the location of Markarian 55.
}
\end{figure}

The $Chandra$ ACIS-I observations were centered at ($\alpha$,
$\delta)=(194.50226^{\circ}$,$27.36541^{\circ}$), 41.4 arcminutes 
from the center of the cluster, which is defined as the center of 
cD galaxy NGC 4874 [located at ($\alpha$, $\delta)= (194.89875^{\circ}$
,$27.959167^{\circ}$)].  
The $\approx256$~arcmin$^{2}$ ACIS-I field of view
corresponds to $\approx0.22$~Mpc$^{2}$ at the distance of Coma. 
The X-ray observation, 65 ks in length (of which 60 ks was usable), was
obtained during 2004 March and cover roughly 20\% of the area covered
optically by \cite{Mobasher2001spectra}.  
The field was chosen as a compromise between 
galaxy density and ICM brightness while remaining
within the optical survey boundaries 
(see Figure~\ref{big_image} for the $Chandra$ observation location).

Our basic data reduction includes removal of pixel randomization and 
correction for the charge transfer inefficiency
created by radiation damage to the detector \citep{Townsley02}.   
We used the routine provided by the $Chandra$ X-ray Center (CXC), 
{\sc wavdetect}, 
to search for X-ray sources down to a relatively low threshold of 
$1\times10^{-7}$ (this is the probability that the source is false) 
in the 0.5--8~keV (full), 0.5--2~keV (soft) and 2--8~keV (hard) X-ray bands.  
We then used the {\sc acisextract} (v3.2) tool 
\citep[developed at Penn State;][]{acisextract},
 to perform additional reductions including constructing PSFs 
and calculating weighted RMFs/ARFs. A total of 74 X-ray point sources were
detected in this field.    We also ran source-searching at a much lower
X-ray detection threshold to aid in placing X-ray upper limits on undetected
sources ($1\times10^{-4}$ threshold).

\begin{figure}[t!]
\centerline{\includegraphics[scale=1.00,angle=0]{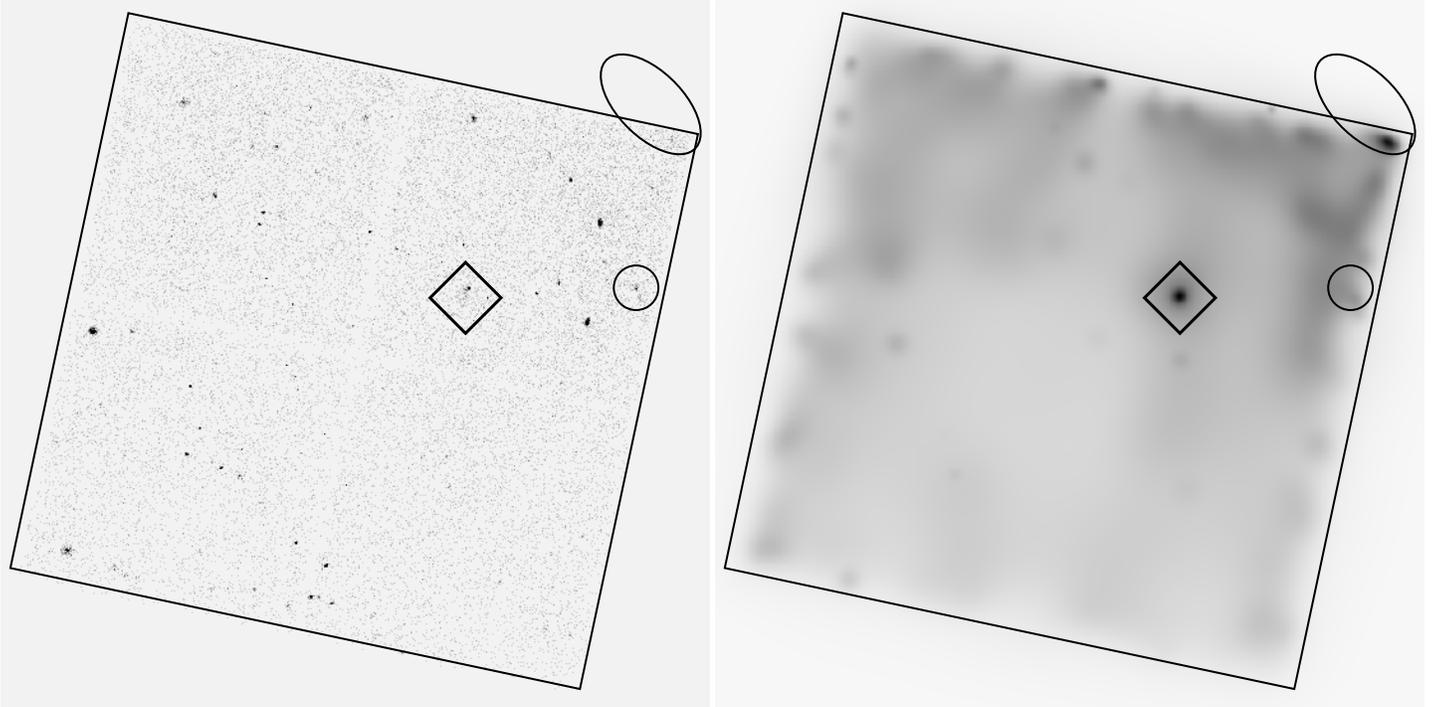}}
\caption[]{ \label{xray_image} {(\bf RIGHT:)} Raw 0.5--2~keV $Chandra$ image
 (binned by 4 for ease of viewing), centered at $\alpha=194.5022$,$\delta=27.3654$.   
The large ellipse and small circle mark NGC~4839 and Markarian 44 as in Figure~\ref{big_image}.
{(\bf LEFT:)} Smoothed 0.5--2~keV $Chandra$ image of the COMA3 region 
(the 74 X-ray point sources have been removed).  
Note that the background is higher to the North and that the NGC~4839 group is off-field
towards the north.  A diamond of side length 1$^{\prime}.5$ marks the location of a possible region of diffuse
X-ray emission, that appears very close to two $Chandra$ point sources to the North (shown in more 
detail in Figure~\ref{diffuse_figure}).} 
\end{figure}

We also search for diffuse structure in order to characterize the ICM in 
the field (although a detailed analysis of the cluster ICM emission is 
beyond the scope of the present study).   We removed regions corresponding 
to the 90\% encircled-energy radius of the PSF at each source's location 
and ran the $csmooth$ routine over the resulting image 
(see Figure~\ref{xray_image}).   There is clearly a background gradient in the
field with brighter ICM emission to the North \citep[this is the direction 
of the NGC~4839 group, which is thought to be falling into Coma
and where the secondary peak of the cluster emission is located; ]
[]{Neumann2001}. There was one particularly bright region in the northwest 
corner of our field, which is clearly visible both in the smoothed and
the raw $Chandra$ data (this is circled in Figure~\ref{xray_image}).  
There are two X-ray point sources to the north of this region.  Due to 
the sub-arcsecond imaging of $Chandra$, we may 
safely reject any contamination from either of these X-ray point sources 
to this region of diffuse X-ray emission.  We discuss this diffuse region in
more detail in Section~\ref{diffuse}.

We used the resulting background map to determine the X-ray sensitivity, 
following the algorithm of \cite{davocatalog}.  A background map was 
constructed by removing regions with radii containing 90\%
encircled-energy for the 74 point sources. The 3$\sigma$ sensitivity was 
then determined by measuring the remaining background at each position in 
the ACIS field using Equation 2 of \cite{Muno2003} and then calculating the
sensitivity based upon a limiting signal-to-noise ratio (we chose 4$\sigma$ ). 
We also measured the background directly in larger regions and found a general
increasing gradient from the southeastern corner (where the background is $\approx0.02$ counts/pixel
in the 0.5--8~keV band) to the northwestern corner, near NGC~4389, where the background
is $\approx0.04$~counts/pixel.

\begin{figure}[t!]
\centerline{\includegraphics[scale=0.80,angle=0]{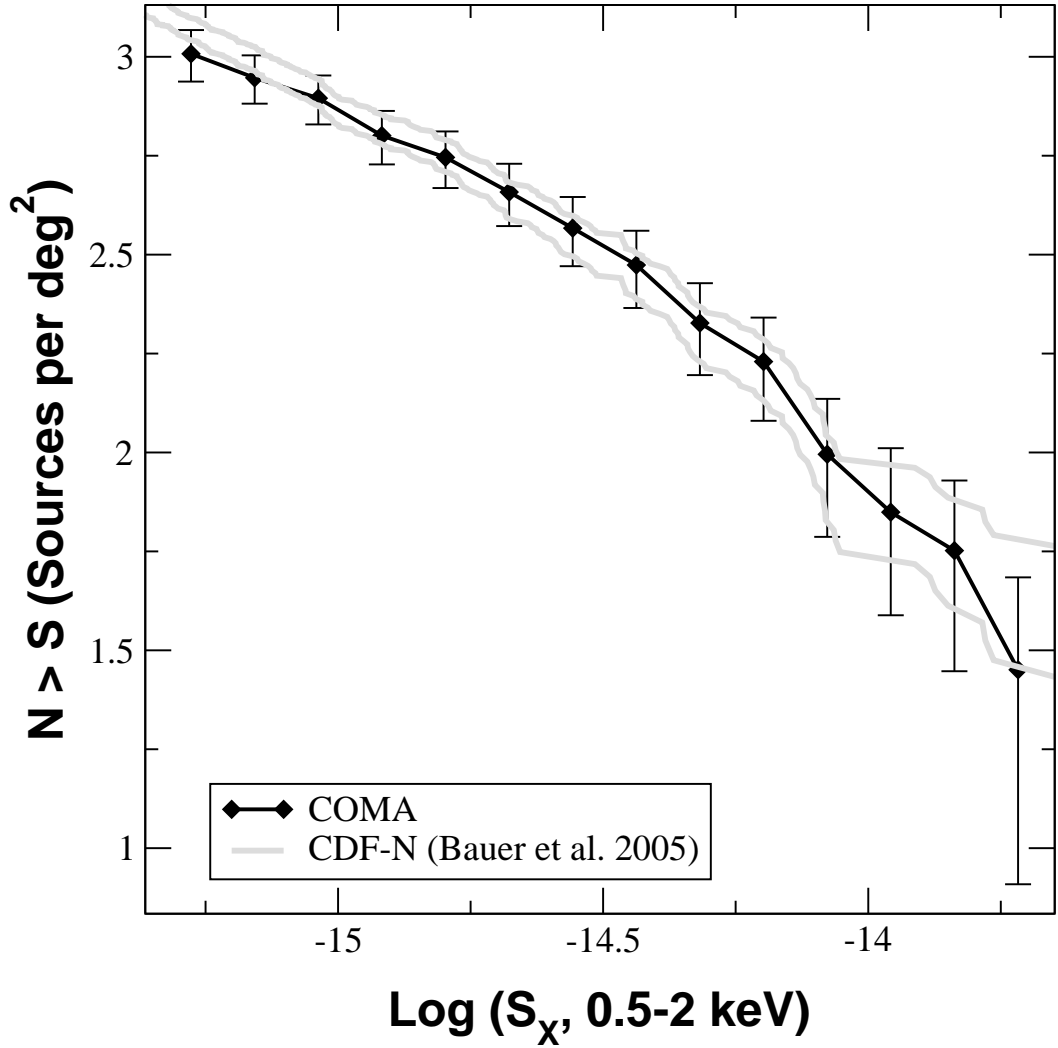}}
\caption[]{\label{LogNLogS} 0.5--2~keV X-ray number counts in the $Chandra$ field
compared with the expected background number counts from \cite{BauerLogN}.  
Statistically, most of the X-ray sources in the field are background AGN.  } 
\end{figure}

We concentrate the analysis in this paper on the 0.5--2~keV band due to the
greater effective area and overall sensitivity to point sources in the soft
band.  Using the X-ray sensitivity map and limiting our
analysis to the 62 sources detected with at least 10 0.5--2~keV counts, 
we constructed the $\log N$-$\log S$ plot for the $Chandra$ survey to 
determine how many of the X-ray sources are likely to be background
AGNs. The number counts here were compared with those in \cite{BauerLogN} 
for the 2~Ms CDF-N (see Figure~\ref{LogNLogS}).  

Clear statistical excesses of X-ray sources have been found in many different
$Chandra$ surveys of galaxy clusters \citep{Cappi01, Sun2002, Cappe05, Ruderman2005}. 
Generally, these other {\it Chandra}-observed clusters are much farther away than Coma so the coverage is
wider-field and the typical detected sources are more luminous (they are
largely AGN).   In contrast our field is small and the coverage is deep.  
The Coma number counts demonstrate that, 
{\it statistically}, the X-ray sources in this field are consistent with the 
expected background AGN population.  Based on cosmic variance alone 
\citep[$\sim30$\% among the deepest $Chandra$ surveys;][]{BauerLogN},
it is still possible that up to 20 of the X-ray sources in the field
could be Coma-member galaxies.  Our results are consistent with
the  \cite{Fino2004} Coma number counts analysis over a wider field 
(that most of the X-ray sources are background AGN).  Although we are
not able to {\it directly} measure the Coma member number counts in this
field, we can constrain their numbers using spectroscopic identification
(see \S~\ref{efficiency} for more details).

\section{Optical Photometric and Spectroscopic Data }
\label{opticalsection}

The Coma cluster has been the target of a number of redshift surveys.  
An extensive study was recently completed with the William Herschel
Telescope (WHT), comprising wide-area
imaging in the $B$ and $R$ bands \citep{Komiyama2002} over multiple fields
and medium resolution spectroscopy (Mobasher et al 2001) in two fields:
one at the center of Coma and one at an off-center location.
Each imaging field covers an area of 
$\approx 30\times 50$~arcmin$^2$ and is photometrically complete to 
$R\sim 21$ mag. The X-ray observations presented here
cover a subset of the off-center spectroscopic field, which is
located at a distance of $\sim40^{\prime}$ from the center
($\alpha=12^{\rm h}59^{\rm m}42.8^{\rm s}$,
$\delta=+27^{\circ}58^{\prime}14^{\prime \prime}$; see Figure~1). 

In this study we use the redshifts from Mobasher et al. (2001) 
together with the photometric data from the wide-area Coma survey.
The redshift survey of Mobasher et al. (2001) was not spectroscopically 
complete due to the high number density of background galaxies.  A color
cut-off of $B-R<2$, shown to be effective in rejecting background galaxies was 
imposed.  There are two galaxy samples, a bright sample (corresponding to
giant galaxies) and a faint sample (corresponding to dwarf galaxies).  Both were
selected as a random subset (checking to make sure the color-magnitude distribution
matched the parent galaxy sample) for spectroscopic follow-up for $R<18$ ($M_R\approx -17$) and $18 < R < 20$ ($-17 < M_R < -15$).    
Thus, while there is a robust measurement of the optical luminosity 
function in this region,
not every galaxy has a spectroscopic identification.   For the purposes of
this paper, we define a split between giant and dwarf galaxies at $M_R=-18$. 

Note that while the optical photometry and spectroscopy in this field is
sufficiently deep to detect relatively faint dwarf galaxies at the distance 
of Coma ($M_R=-14$), the limits are not faint enough to obtain redshifts
for all of the $Chandra$ background AGN.   However, wider-field work by
P. Martini et al. 2006, in preparation on a sample of 8 clusters at $z<0.3$
(all are more distant than Coma) shows that the spectroscopically identified
sources do account for the statistical excess.  Thus, we can be fairly sure
that we are not missing Coma-member galaxies in the X-ray source population
lacking optical counterparts.

\section{$Chandra$ Sources with Optical Counterparts}
\label{optcounterparts}

\begin{figure}[t!]
\centerline{\includegraphics[scale=0.80,angle=0]{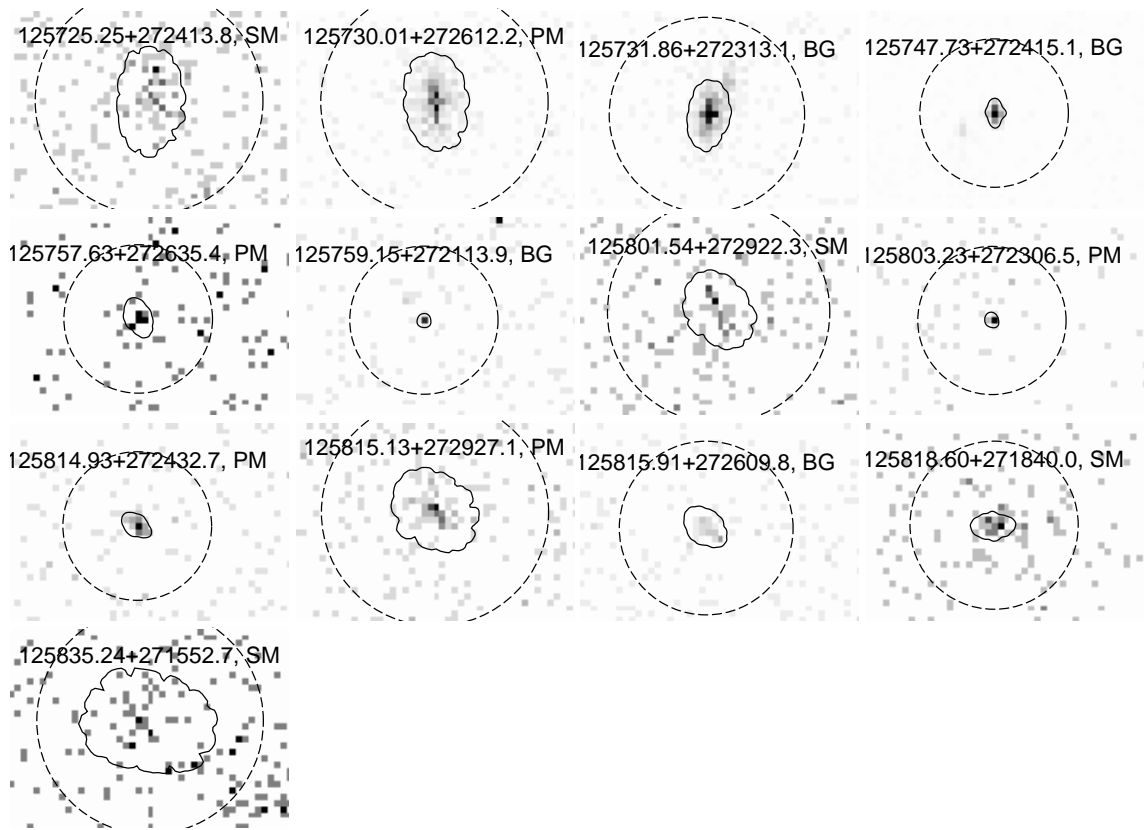}}
\caption[]{\label{xray_postagestamps} Raw 0.5--8~keV $Chandra$ images of the sources in
Coma.  The inner contours show the 90\% encircled energy contours calculated 
for each source location with {\sc acis\_extract} and the outer circles 
show the background annulus used for each region.   The labels for each source indicate Spectroscopic Member (SM), Photometric Member (PM), and likely Background AGN (BG).   The photometric 
membership is based upon optical colors, discussed later in \S~\ref{opticalsection}.}
\end{figure}

Table~1 presents the 13 $Chandra$ X-ray 
sources with optical counterparts above
the $R=21$ limit of the WHT imaging survey.  Figure~\ref{xray_postagestamps} shows the
raw 0.5--8~keV X-ray images for these 13 sources.   Clearly the vast majority of the
$Chandra$ sources with bright optical counterparts are consistent with point-like
X-ray emission.  The spectroscopically-confirmed Coma-members (whose angular
extent is also the largest) are less obviously point-like and may be moderately extended.

\begin{figure}[t!]
\centerline{\includegraphics[scale=0.70,angle=270]{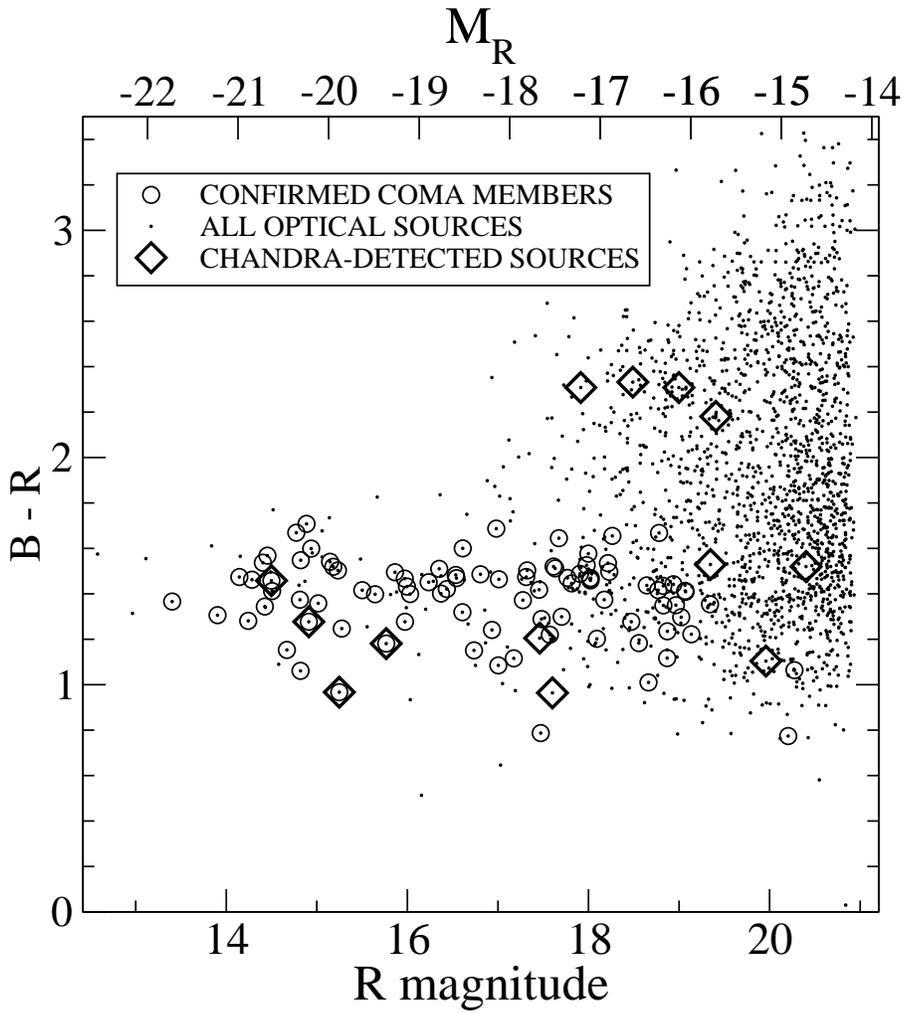}}
\caption[]{\label{colors} $B-R$ optical color vs $R$ magnitude for galaxies in the
Mobasher et al. (2001) survey.  The dots mark all photometrically-detected sources
whereas the circles indicate those sources with spectroscopically-confirmed Coma membership.
Four of the X-ray source are in the region of this plot that is mainly populated by
background sources.

\label{postagestamps}}
\end{figure}

\begin{figure}[t!]
\centerline{\includegraphics[scale=0.70,angle=0]{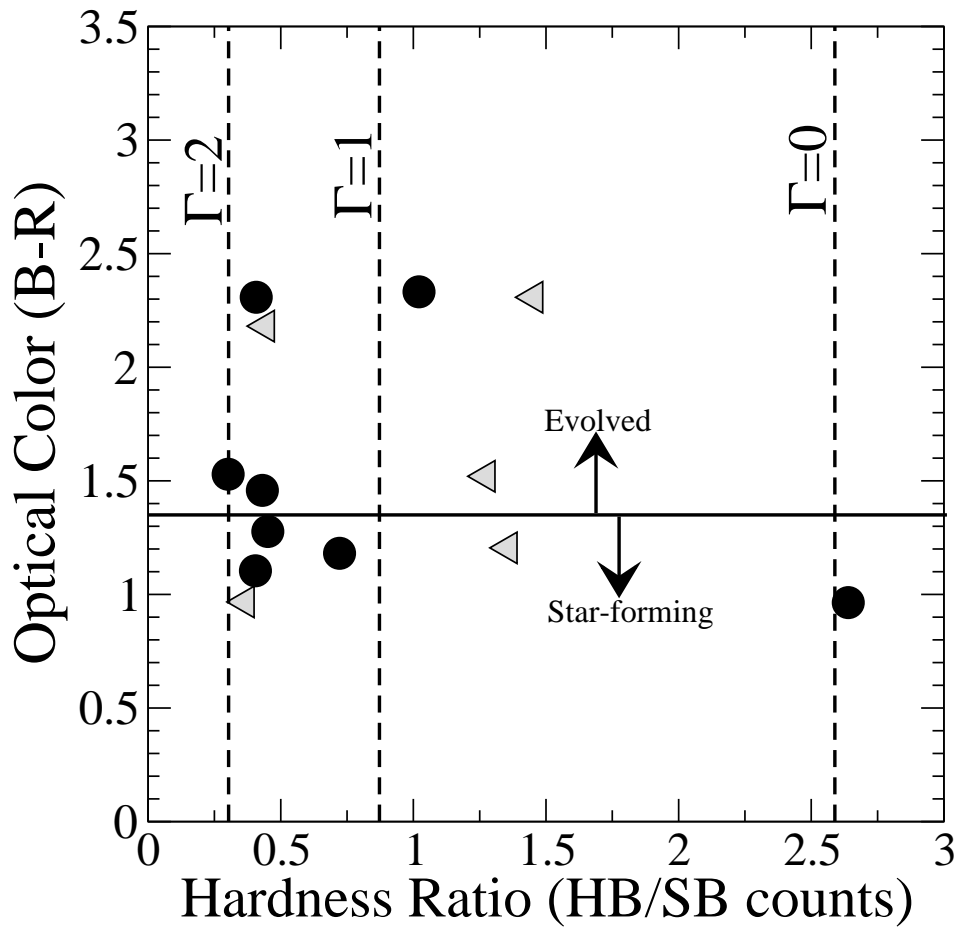}}
\caption[]{Optical $B-R$ color versus X-ray hardness ratio for all $Chandra$ sources with
optical counterparts brighter than $R\approx21$.  The triangles indicate upper limits (99\% 
confidence on the hard X-ray detections).   Among sources with $B-R<2$, indicative of
cluster membership, there is a small amount of preference for lower hardness ratios, indicating
the sources are not as X-ray hard.  This favors the interpretation that these sources are
dominated by non-AGN emission (see text).  The X-ray hard source in the ``star-forming" portion
of the plot is discussed in more detail in \S~\ref{optcounterparts}.  \label{xray_vs_color} 
}

\end{figure}

Figure~\ref{colors} shows the locations of the Chandra sources with
photometric matches on a color-magnitude plot ($R$ vs $B-R$).
The spectroscopically confirmed Coma
members are also indicated (all have $B-R<2$ due to the target selection; 
see \S~\ref{opticalsection}).
Four of the X-ray sources have colors that are
redder than typical Coma-member galaxy colors, indicating they
are likely background AGN.   In Figure~\ref{xray_vs_color}, 
we plot the $B-R$ color
versus X-ray hardness ratio (plotted as a ratio of 2--8~keV/0.5--2~keV counts).  
Note that only 8 of the 13 sources had confident detections in the 
2--8~keV band.   Among the sources with optical colors indicative of cluster membership
($B-R<2$), there is some preference for lower hardness ratios, indicating that these sources
are likely not obscured AGN (note that none of these sources contain unobscured AGN; see
\S~\ref{spectroscopic_members}).   
We have also overlaid the color cut suggested by Mobasher et al. (2001) and verified using
ultraviolet observations by D. Hammer et al. 2005, in preparation, for a division 
between evolved and star-forming galaxies.   There is little difference in X-ray hardness 
across this line.  However, the hardest X-ray source is in the star-forming region of the plot
(CXOCOMA~J123815.1+272927).  Its 2--8~keV luminosity is $\approx8\times10^{39}$~\lumin 
if it is indeed a Coma member.  It is quite possible that an individual ULX might be 
dominating the X-ray emission from this galaxy, indeed they can be quite X-ray hard
\citep[e.g., see][]{HornULX}.  The angular offset is marginal, even with {\it Chandra}, however.
Since there is no spectroscopic redshift, we don't comment on this source any further.

\begin{figure}[t!]
\centerline{\includegraphics[scale=0.70,angle=270]{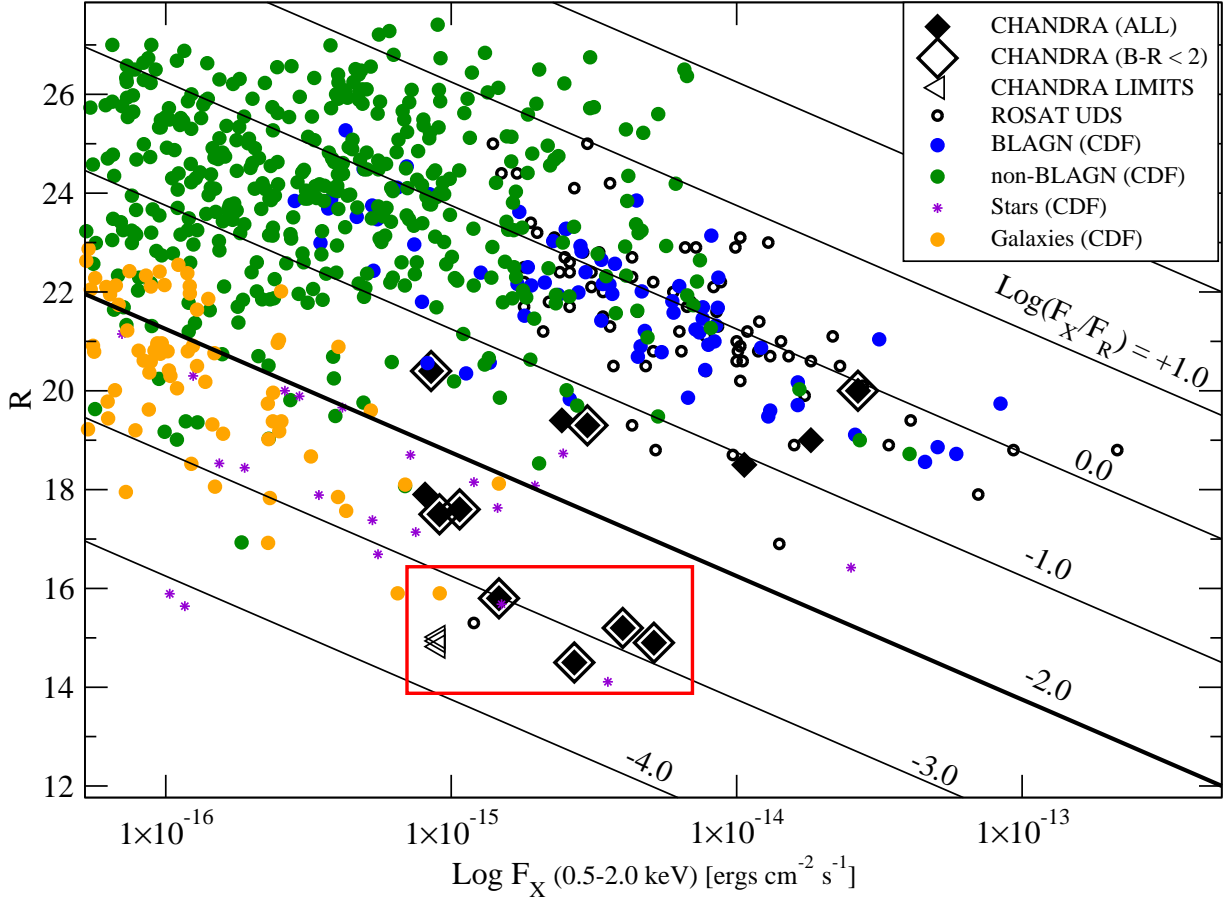}}
\caption[]{ \label{RFx} $R$ vs f$_{X}$ for the 13 Chandra-detected Coma-area galaxies and the three
undetected giant galaxies (triangles).  For comparison,
we show the $ROSAT$ UDS data of \cite{Lehmann01} and Chandra Deep Field (CDF) data on Broad-Line AGN (BLAGN), non-Broad Line AGN (mostly narrow-line
AGN), stars, and ``normal" galaxies \cite{BauerLogN}.   Most of the X-ray detected Coma-area sources are 
clearly in the normal galaxy part of the plot.   The red box shows the location of all the spectroscopically
identified Coma members, all of which are consistent with a low X-ray/optical flux ratio.
}
\end{figure}

In Figure~\ref{RFx}, we show the $R$-band vs X-ray flux plot for these 
same 13 sources, with the sources having
$B-R<2$ marked.   This figure demonstrates the success in reaching low X-ray-to-optical
flux ratios in a moderate-depth $Chandra$ observation of Coma.   In total, there are six sources 
that have both $B-R<2$ colors indicating Coma membership and
the lower X-ray-to-optical flux ratios indicating that they are likely to
be normal galaxies (i.e. are not AGNs).  All six 
sources are brighter than $R=18$, corresponding to the optical survey's giant
galaxy optical cutoff.   We also show upper limits for the three {\it undetected}
Coma-member giant, spectroscopically-confirmed, Coma-member galaxies in the field.


\section{X-ray Constraints on Spectroscopically-Confirmed Coma Members}
\label{spectroscopic_members}

There are a total of 23 spectroscopically-confirmed $Coma$-member galaxies
from the \cite{Mobasher2001spectra} survey within the $Chandra$ field.  
Of these, 7 are giant galaxies and 16 are dwarf galaxies. 
Four of the seven giant galaxies are detected by $Chandra$ 
and none of the dwarf galaxies are detected (see Table~1).   
We have plotted the 7 undetected giant galaxies
as X-ray upper limits in Figure~\ref{RFx}.  

\begin{figure}[t!]
\centerline{\includegraphics[scale=0.70,angle=0]{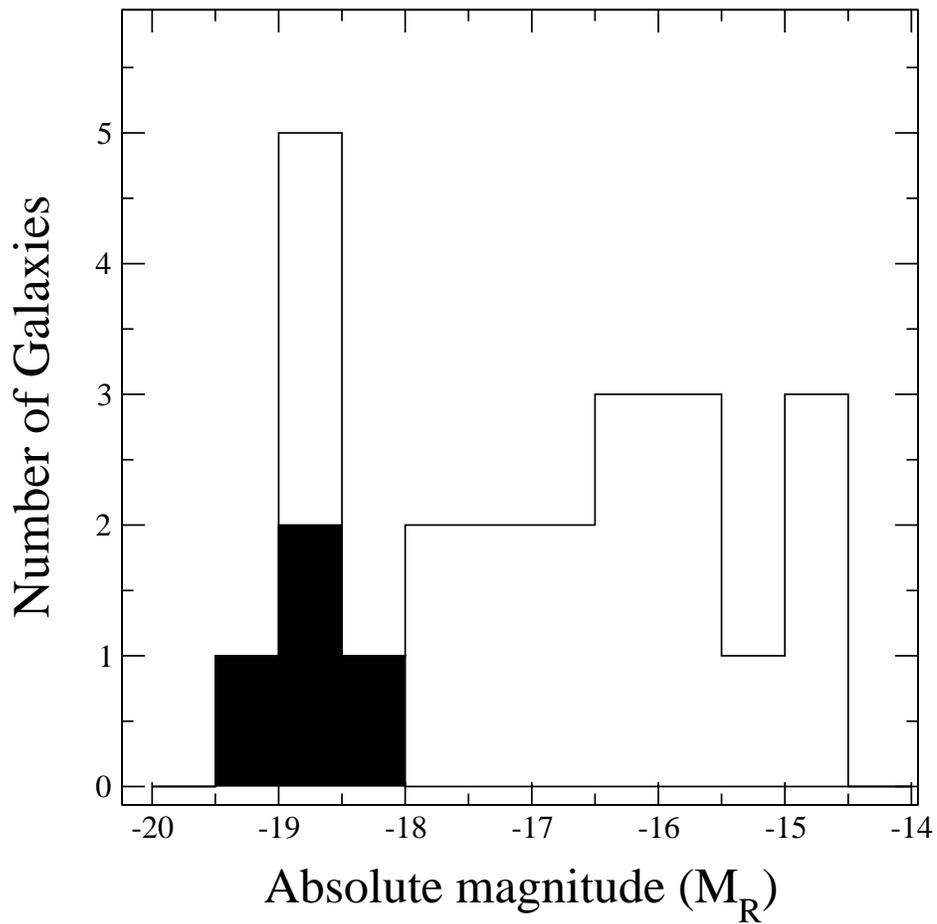}}
\caption[]{\label{mags} The histogram shows the absolute $R$-band magnitudes for spectroscopically 
confirmed Coma members in the $Chandra$ field with the filled-in portion of the histogram
showing the $Chandra$-detected galaxies. 
}
\end{figure}

To place the detections and non-detections into better physical context,
we have plotted the {\it absolute} $R$-band magnitudes of all 23
galaxies in Figure~\ref{mags}. 
We checked our lower-significance X-ray catalog to
determine if we can detect {\it any} X-ray emission from
the three optically bright X-ray-undetected galaxies.
 We also tried running {\sc wavdetect} with
different wavelet scales (for instance, neglecting the small scales and
adding larger scales to try to detect spatial extent).  Our sensitivity
calculations indicate that our {\it worst} sensitivity even for a moderately
extended source (over several arcseconds) is $<2.6\times10^{-15}$~\flux
(0.5--2~keV) for all three sources, corresponding to
X-ray/optical flux ratios of $\log{f_{X}\over{f_{R}}}<10^{-3}$--$10^{-3.2}$.

\begin{figure}[t!]
\centerline{\includegraphics[scale=0.70,angle=0]{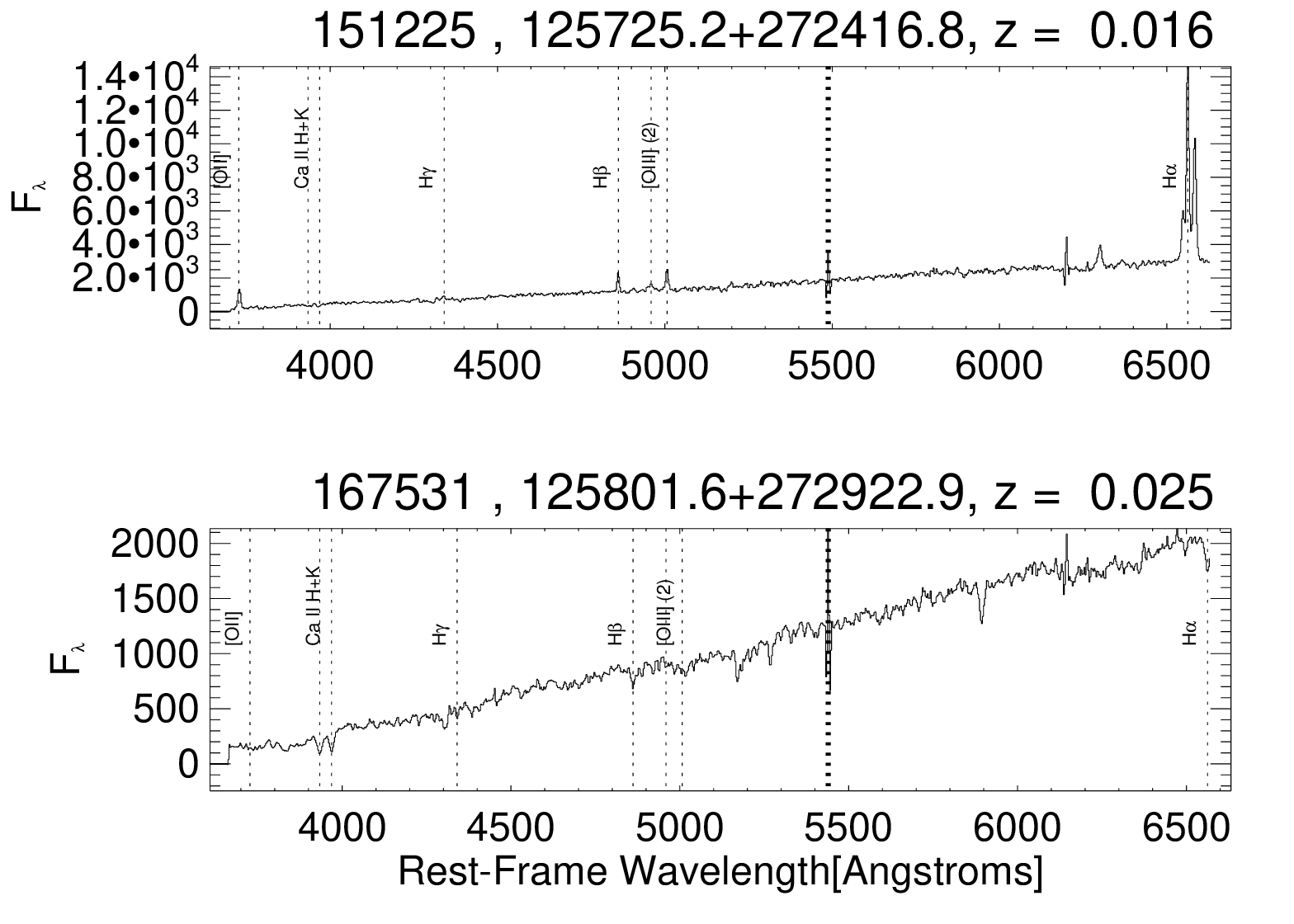}}
\centerline{\includegraphics[scale=0.70,angle=0]{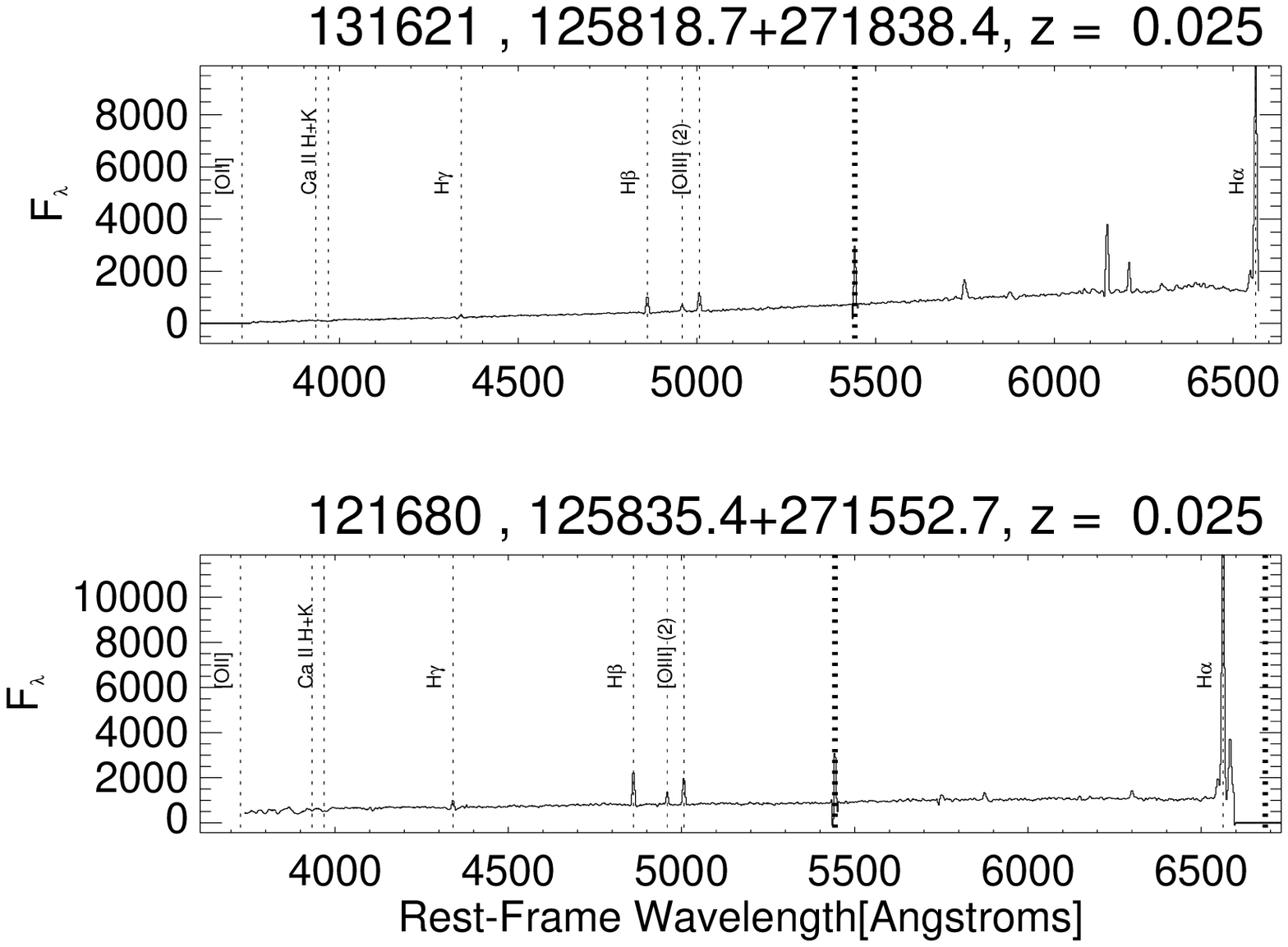}}
\caption[]{\label{spectra_detected} 
Optical spectra of $Chandra$-detected galaxies from the 
work of \cite{Pogg2001spectra} in
rest-frame wavelengths.  The identification number on each spectrum is
the index from the main photometric catalog of \cite{Mobasher2001spectra}.  
The locations of several key spectral diagnostic lines have been 
marked on all four spectra.  The wide, dashed region
indicates the location of a telluric OI emission line (5577~\AA).  
}
\end{figure}

\begin{figure}[t!]
\centerline{\includegraphics[scale=0.70,angle=0]{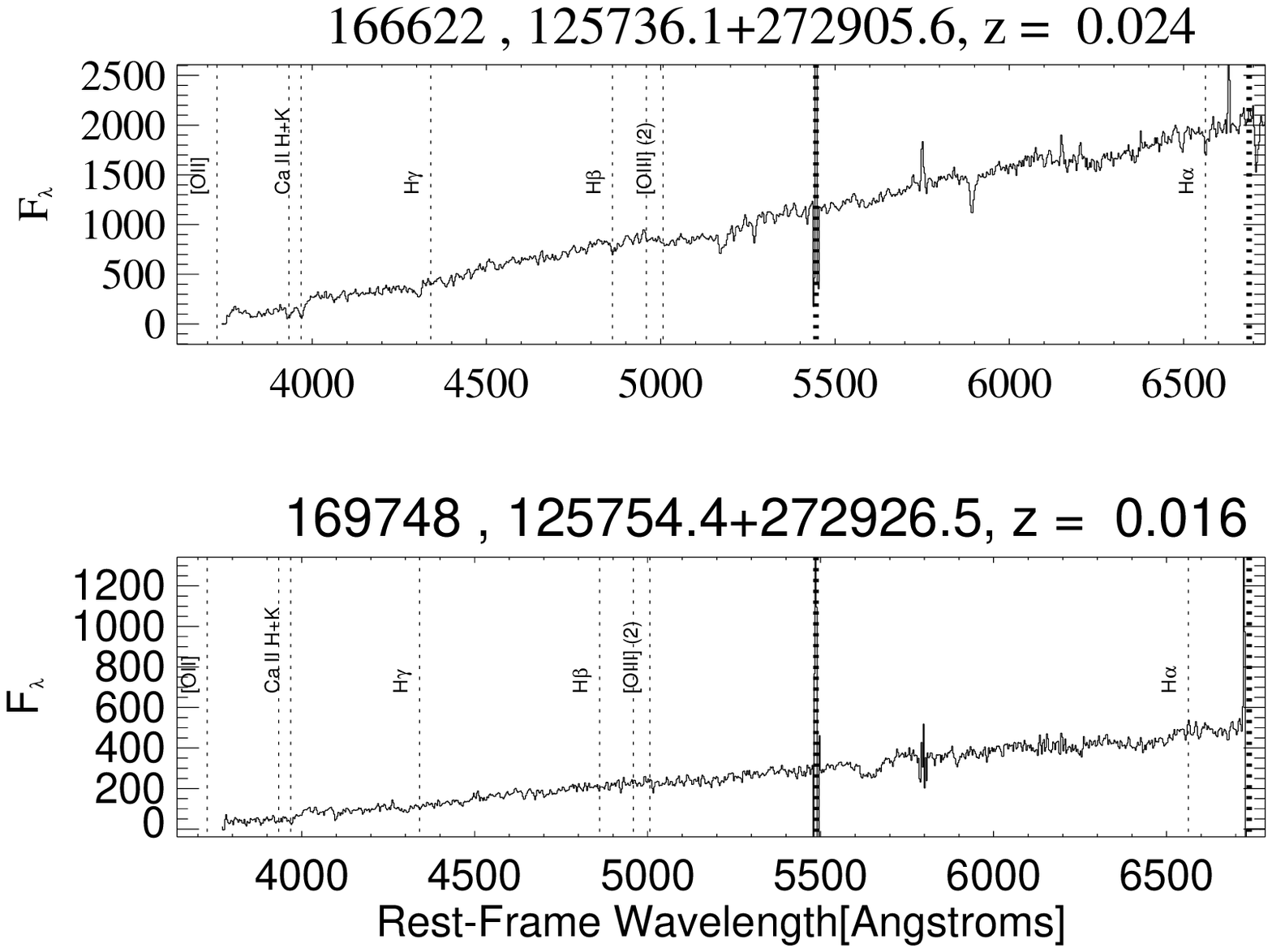}}
\centerline{\includegraphics[scale=0.70,angle=0]{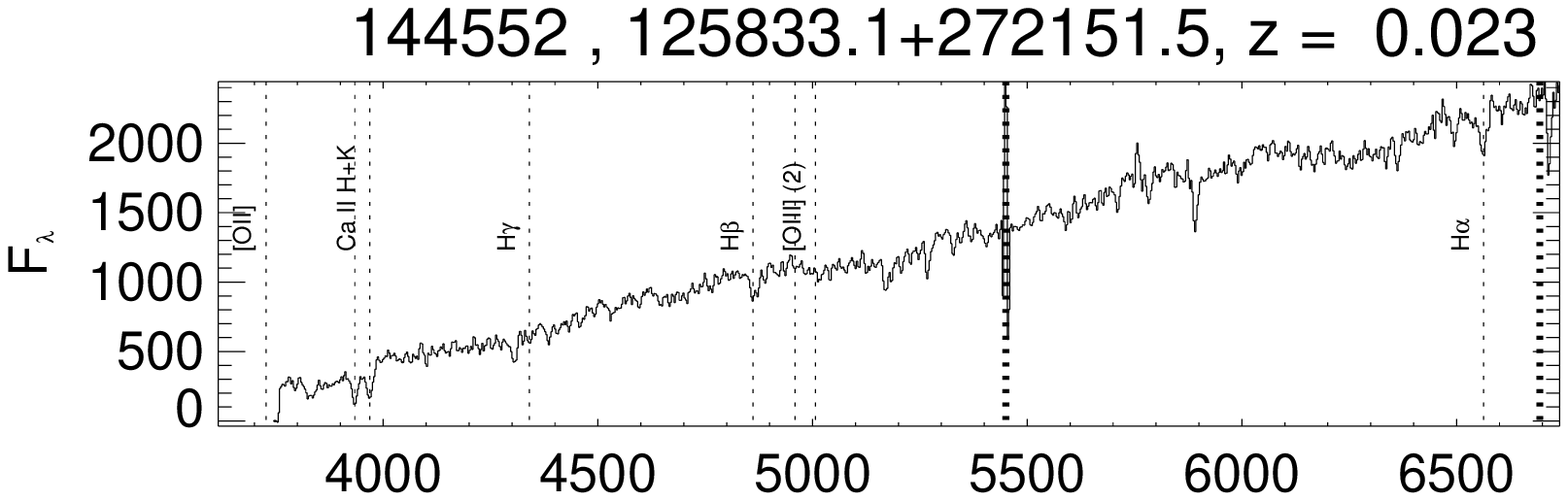}}
\caption[]{\label{spectra_undetected} Same as Figure~\ref{spectra_detected} except these are the three
spectroscopically-confirmed giant Coma-member galaxies without X-ray detections.}
\end{figure}

Figures~\ref{spectra_detected} and \ref{spectra_undetected} show the optical
spectra of the 4 X-ray-detected and 3 X-ray-undetected giant galaxies.  Three of the
four X-ray-detected sources are emission-line galaxies indicative of AGN and/or 
starburst activity whereas the undetected galaxies are mostly absorption-dominated,
``passive" systems.    The one X-ray detected galaxy with an absorption-dominated
optical spectrum has the lowest X-ray/optical flux ratio (10$^{-3.3}$).   This clearly
indicates that the three X-ray undetected optical giant galaxies may just be narrowly
below the sensitivity of this survey.

We have measured the emission lines to further characterize their nature, comparing with 
the diagnostic lines of \cite{Veill87}.  We considered the emission line ratios
OIII/H$\beta$ vs OI/H$\alpha$ and  OIII/H$\beta$ vs NII/H$\alpha$.  Of the three emission-line
galaxies (see Table~2), one is clearly in the HII galaxy (star forming galaxy) region of these
diagnostic plots.   For the other two, we do not have good H$\alpha$ coverage or it is ambiguous:
it could be either a LINER or an HII galaxy.  Neither are consistent with Seyfert galaxies, however.
All of the absorption line (k-type galaxy) spectra are ``normal" (passively evolving, nothing
special) except for the 125754.3+272926 which is a k+a (possible post-starburst) galaxy.  This source
was not detected.

\section {Efficiency of Detecting Galaxies in the X-rays in Coma}
\label{efficiency}

Once we have identified the $Chandra$ sources that are likely Coma members, we may
estimate the ``efficiency" of detecting galaxies in the X-ray band in this field versus
a blank field survey.  If we pessimistically assume that we have detected only four Coma-member
galaxies in the X-ray band that are Coma members, this corresponds to $\approx50^{+37}_{-25}$ sources deg$^{-2}$ (statistical errors, 90\% confidence)  at our flux limit of $\approx2\times10^{-15}$~\flux (0.5--2 keV).   Note that all
four of these galaxies have {\it low} X-ray/optical flux ratios (see Figure~\ref{RFx} ) and have optical spectra
that support that they do not have significant contributions due to hidden obscured AGN (see \S~\ref{spectroscopic_members}).
We consider
three studies of field galaxy number counts in the X-ray band.   The deepest survey, the
2~Ms CDF-N has the highest galaxy number density detected thus
far, down to X-ray fluxes of $\approx3\times 10^{-17}$~\flux \citep[0.5--2 keV; ][]{BauerLogN}.
Note that the \cite{BauerLogN} study {\it does} include low-luminosity AGN, as do all
such field galaxy surveys.  Indeed, since most galaxies exhibit lower-level AGN activity
\citep{Ho01,Ho2003}, it is not really possible to completely separate the X-ray sources 
into two populations (galaxies and AGN).  \cite{BauerLogN} obtained source counts 
of $\approx1.9$ and $\approx3.6$ galaxies per square degree at
a 0.5--2~keV flux $\approx2\times10^{-15}$~\flux ~depending on how aggressively one excludes possible
moderate-luminosity AGN.
 We also compare with the \cite{Georgakakis04} {\it XMM-Newton}
Needles in the Haystack survey at $z\approx0.1$ which covers intermediate X-ray fluxes more appropriate to those
here.  The \cite{Georgakakis04} estimate also includes some AGN contamination, at $\approx2\times10^{-15}$ (0.5--2~keV)
their number counts are $\approx7$ galaxies deg$^{-2}$.  It is difficult to compare
with a truly {\it local} ($z<0.1$) estimate of the number counts; the best
such estimate appears to be the the {\it ROSAT} study  of \cite{Tajer05} which
reaches $\approx10^{-14}$~\flux (0.5--2~keV) but includes a higher AGN fraction
than the other studies (many sources have higher values of X-ray/optical
flux ratio and clear indications of AGN activity, although at relatively low
levels).  \cite{Tajer05} finds $\approx18$ galaxies~deg$^{-2}$ at 
$\approx2\times10^{-15}$ (0.5--2~keV).  Due to the high AGN fraction we do not use
the \cite{Tajer05} number in our comparison.  We may thus conclude that this 
Coma survey is 7--26 times more efficient at detecting galaxies if we have just four
Coma members in our $Chandra$ field.   

\section{Constraints on the Normal Galaxy X-ray Luminosity Function in Coma}

Comparison of the cluster XLF with the field XLF is tricky business, as outlined
in \cite{Fino2004}.  One must carefully de-project the distribution of galaxies
over the volume of the cluster to determine the number of galaxies per unit volume.
Within a cluster of galaxies, the LF is generally calculated per unit {\it area} so
such deprojections are not required.  The optical LF with which we compare is measured
per unit {\it area} whereas typical field LFs are per unit volume.  We do not do a 
field/cluster comparison as we do not directly measure the LF.

\begin{figure}[t!]
\centerline{\includegraphics[scale=0.80,angle=0]{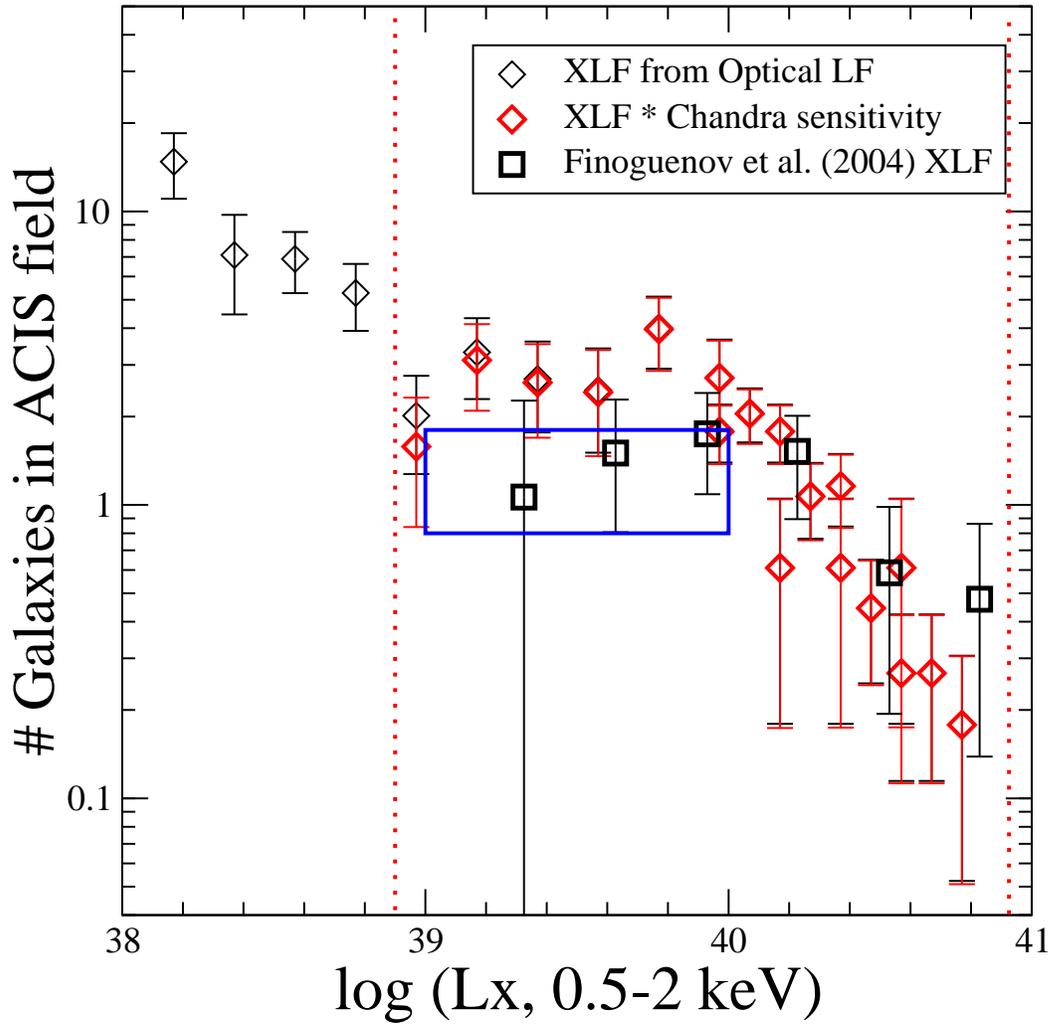}}
\caption[]{\label{XLF} Converted optical LF, assumed $\log{f_{\rm X}\over{f_{R}}}=-3$ (typical of the
Milky Way galaxy and many galaxies in the nearby Universe, e.g. Shapley et al. 2001).

\nocite{Shapley01} }

\end{figure}

We have taken the well-measured optical luminosity functions of 
Mobasher et al. (2003; from their Table 3) and converted them to X-ray luminosity functions for
comparison with this survey.  We are encouraged by the apparent success of
translating non-X-ray (generally infrared) LFs to XLFs 
from other studies \citep{Norman2004,Ranalli2005}.   
%
%
We converted from optical magnitudes to X-ray fluxes 
assuming $\log{f_{\rm X}\over{f_{R}}}=-3$, the value favored by all the
spectroscopically identified sources, and which also appears typical of e.g.,
the Milky Way.
It appears from the non-detections that
even lower values of the X-ray-to-optical flux ratio may be plausible, 
but we choose a conservative scenario for the current comparison.
We also plot this translated XLF with corrections for the sensitivity derived in \S~\ref{chandraobs} to determine the range
of the XLF probed by our Chandra observations.   The red data
points in Figure~\ref{XLF} shows the expected $Chandra$ XLF corrected for
the area of the field where we are able to detect a galaxy of the given
luminosity.   We find that our sensitivity drops sharply below
$\approx10^{39}$~\lumin (0.5--2~keV).
To check to see how reasonable this assumption of a constant 
$\log{f_{\rm X}\over{f_{R}}}=-3$ might be, we compare with the \cite{Fino2004} 
X-ray luminosity function (which was determined over a much larger area in 
Coma).   Since the \cite{Fino2004} $XMM$ XLF is primarily for the center of the cluster, we
scale it down by a factor of 0.61 to account for the reduced galaxy density in our chosen
field.  

We find that our XLF, which is just the optical LF for this field translated assuming
$\log{f_{\rm X}\over{f_{R}}}=-3$, agrees well with the Finoguenov XLF, which is a 
{\it directly} measured XLF.  We can look at the {\it number} of X-ray detected galaxies
in our sample and compare it with the predictions of this XLF.  If we integrate the
converted XLF from $10^{39.2}$--$10^{40.2}$~\lumin , which corresponds to our 
approximate detection limit on the faint end, we find that $\approx14$ galaxies
are predicted in this field.  
We
use the total number of X-ray detected galaxies (the upper bound corresponds to 9 members whereas the
lower bound is the spectroscopically-confirmed subset) 
 as a constraint on the Chandra XLF (shown as a blue box in Figure~\ref{XLF}).
We find good agreement between our constraint (which is admittedly broad)
and the Finoguenov XLF.   Both XLFs thus support a very low value of X-ray-to-optical flux ratio
(less than $10^{-3}$).  

However, our results are obtained in an off-center field (largely outside of the region
studied by Finoguenov). This might indicate that all of the
impact of the cluster ICM on X-ray/optical flux ratios occurs at relatively low ram
pressures at 1.1~Mpc from the cluster core.  This might be expected based on the
cluster ICM interactions observed by \cite{Neumann2001}, which showed that the Coma ICM
at this radius exerts enough pressure to displace the intragroup medium of the 
NGC~4839 group.  The potential well of a group is of course deeper than that of galaxies, so ICM
effects are likely still expected.     This points to galaxy infall as a major contributor
to the properties of galaxies in clusters.   Indeed, recent optical surveys are finding
that star-forming galaxies are found to be more prominent in infalling structures in 
clusters at higher-redshift \citep[$0.3 < z < 1.0$; e.g., ][]{Tran2005}.  The galaxies
that we are studying in Coma, particularly in an area relatively nearby the NGC~4839 group,
are the most likely descendants of these star-forming galaxies.

\section{Diffuse X-ray Emission in the COMA3 field}
\label{diffuse}

As described in Section~\ref{chandraobs}, we identified a region of diffuse X-ray
flux in the $Chandra$ field.  We constructed a 20$^{\prime \prime}\times30^{\prime \prime}$ 
region centered on the peak of the X-ray flux in the adaptively smoothed image,  
and a background frame, starting 5 arcseconds outside this box, and ending 15 arcseconds 
outside the box.   There were 57 0.5--8~keV X-ray counts within this box-shaped region, 
39.9 above background corresponding to flux of
$6.9 \times10^{-15}$~erg~cm$^{-2}$s$^{-1}$ (0.5--8~keV), 
or $\approx 9.3\times10^{39}$~\lumin at the distance of Coma.
Comparison with the X-ray luminosities of known galaxy groups in the nearby Universe 
\citep[Figure~7 of ][]{Mulchaey03} shows that this X-ray luminosity is typical 
of lower-luminosity groups.    At this time, we are not able to rule out the possibility
that this is either a foreground or background region of diffuse X-ray flux.

We show the raw X-ray image (with the point sources included) in 
the second panel of Figure~\ref{xray_image}.
The brighter X-ray source is clearly coincident with an optically bright galaxy
 ($\alpha=194.448952$,$\delta=27.404253$, $R\approx 18.5$; $B-R=2.3$) which we show
in the WHT optical image in the third panel.  This source unfortunately did not have
a redshift in the \cite{Mobasher2003LF} catalog and its optical colors indicate that it
is likely a background source unrelated to the diffuse emission.  Spectroscopic observations  
of this galaxy were carried out at the Lick Observatory
during May 2005 but due to poor weather were not successful.
The X-ray flux of $1.0\times10^{-14}$~erg~cm$^{-2}$~s$^{-1}$ (0.5--2~keV) implies an X-ray
luminosity of $\approx1.4\times10^{40}$~\lumin (0.5--2~keV) at the distance of Coma, but it is
more likely a background object.

\section{Conclusion and Future Work}

Through a 60~ks observation in the outskirts of the Coma cluster,  we have 
identified nine X-ray-detected galaxies whose optical colors indicate likely
Coma membership (four with spectroscopic confirmation).  All seven spectroscopically-confirmed
Coma members in this field have detections or limits consistent with very low X-ray/optical
flux ratios.  We have thus confirmed
the suppression of X-ray emission from galaxies in clusters (with respect to their optical
emission) found previously in an XMM-Newton study.  The notable aspect of our result is 
that it is for an off-center region in the cluster.
This may indicate that all of the
impact of the cluster ICM on X-ray/optical flux ratios occurs at relatively low ram
pressures at 1.1~Mpc from the cluster core.   

This particular field would benefit from complete optical spectroscopic coverage to establish
whether any of the $Chandra$ sources with higher X-ray/optical flux ratios might be members (and thus
would run counter to the apparently lower X-ray/optical flux ratios for the current spectroscopic
sample).  These sources {\it could} be moderate luminosity AGN, which would be interesting due to the
relatively high AGN fraction.

Our results indicate that an X-ray survey sensitive to $\approx2\times10^{-15}$~\flux (0.5--2~keV)
can begin to probe normal galaxies at the distance of Coma.  However,
to obtain sufficient numbers of galaxies to directly constrain the XLF 
one would need to observe a larger area (3--4 times the area) or observe to fainter
X-ray fluxes.   Such a wider-field survey would also build upon the existing and planned
excellent multi-wavelength data in the field, including
the recent  Spitzer IRAC (P.I. Hornschemeier) observations which cover
the entire field, as well as the upcoming approved GALEX observations (P.I. Hornschemeier).
Additionally, the Sloan Digital Sky Survey has recently surveyed parts
of Coma; the first portions of these data will be available as part of SDSS DR5.

\acknowledgments

We gratefully acknowledge the helpful comments of an anonymous referee which helped 
clarify the results of this paper.  We gratefully acknowledge the financial support of
\chandra\ X-ray Center grant G05-i5089X,
DMA gratefully acknowledges the generous support from the Royal Society.
We thank A. Finoguenov for sharing data.
 This research has made use of the NASA/IPAC Extragalactic Database which is 
operated by JPL under contract with NASA.
%


\bibliographystyle{apj}
\bibliography{ms}


\clearpage
\thispagestyle{empty}
\begin{deluxetable}{rrrrrcrrrrrrrr}
\rotate 
\tabletypesize{\scriptsize}
\tablewidth{0pt}
\tablecaption{$Chandra$ Sources in the COMA3 Field with Photometric Counterparts}
\tablehead{
\colhead{IAU Name} & \multicolumn{2}{c}{$Chandra$ coordinates} & \colhead{Photo} & \multicolumn{2}{c}{Optical coordinates} & \colhead{R$_{eff}$} & \colhead{$\Delta$XR-Opt} & \colhead{SB flux} & \colhead{HB flux} & \colhead{$R$} & \colhead{$B-R$}  & \colhead{$\log{fx\over{fopt}}$} & NOTES \\
\colhead{} & \multicolumn{2}{c}{} & \colhead{ID$^{\rm a}$} & \multicolumn{2}{c}{} & \colhead{($^{\prime \prime}$)} & \colhead{($^{\prime \prime}$)} & \multicolumn{2}{c}{$10^{-15}$ erg cm$^{-2}$ s$^{-1}$} 
 & \colhead{AB mag}  \\
}
\startdata
125725.25$+$272413.8 &  194.355225 &   27.403835 &  19592 &  194.355211 &   27.404658 &  3.0 &  3.0 & 5.13 & 10.8 &  14.9&  1.3 &  -2.8 & SM \\
125730.01$+$272612.2 &  194.375061 &   27.436747 &  20416 &  194.375144 &   27.436869 &  0.9 &  0.5 &26.60 & 48.6 & 20.0&  1.1 &  -0.1 & PM \\
125731.86$+$272313.1 &   194.382782 &   27.386995 &  18997 &  194.382811 &   27.387107 &  1.2 &  0.4 &18.19 & 37.1 & 19.0&  2.3 &  -0.6 & BG \\
125801.54$+$272922.3 &  194.506424 &   27.489534 &  22196 &  194.506534 &   27.489696 &  5.6 &  0.7 & 2.70 & 4.18 & 14.5&  1.5 &  -3.3 & SM \\ 
125815.13$+$272927.1 &  194.563080 &   27.490862 &  22057 &  194.563418 &   27.490874 &  1.6 &  1.1 & 1.07 & 19.9 & 17.6&  1.0 &  -2.4 & PM, hard$^{\rm d}$\\
125815.91$+$272609.8 &  194.566315 &   27.436064 &  20385 &  194.567152 &   27.436340 &  1.1 &  2.9 & 2.44 & \nodata & 19.4&  2.2 &  -1.4 & BG$^{\rm e}$ \\
  125818.60$+$271840.0 &  194.577515 &   27.311132 &  16898 &  194.577738 &   27.310656 &  3.3 &  1.9 & 1.47 & 3.21 & 15.8&  1.2 &  -3.0 & SM \\
125835.24$+$271552.7 &  194.646866 &   27.264656 &  15769 &  194.647339 &   27.264648 &  3.9 &  1.5 & 3.99 & \nodata  & 15.2&  1.0 &  -2.8 & SM \\
125747.73$+$272415.1 &  194.448883 &   27.404198 &  19518$^{\rm c}$  &  194.448952 &   27.404253 &  1.1 &  0.3 &10.63 & 64.2 & 18.5&  2.3 &  -1.1 & BG \\
  125757.63$+$272635.4 &  194.490143 &   27.443174 &  20668 &  194.490344 &   27.443344 &  1.9 &  0.9 & 0.91 & \nodata & 17.5&  1.2 &  -2.6 & PM \\
  125759.15$+$272113.9  &  194.496460 &   27.353876 &  17981 &  194.496596 &   27.353884 &  1.5 &  0.4 & 0.81 & \nodata & 17.9&  2.3 &  -2.4 & BG  \\
125803.23$+$272306.5 &  194.513489 &   27.385160 &  18934 &  194.513409 &   27.386001 &  1.5 &  3.0 & 0.85$^{\rm b}$ & \nodata & 20.4&  1.5 &  -1.5 & PM \\
125814.93$+$272432.7 &  194.562241 &   27.409094 &  19583 &  194.560907 &   27.408269 &  2.9 &  5.2 & 3.00 & 3.02 & 19.3&  1.5 &  -1.3 & PM \\
\enddata 
\tablenotetext{a}{Photometric identification number from Komiyama et al. (2002)}
\tablenotetext{b}{The flux estimate for this source is highly uncertain as it was observed immediately adjacent to a CCD chip gap.  We estimated the 0.5--2~keV flux from the 0.5--8~keV flux to mitigate the statistical errors.}
\tablenotetext{c}{This is the point source to the north of the diffuse emission.   }
\tablenotetext{d}{This is the ``hard" X-ray source in Figure~6. See \S~\ref{optcounterparts}. }
\tablenotetext{e}{There are two fairly nearby optical counterparts to this source.  This one was closest but another galaxy at 
($\alpha$,$\delta$)=(194.56715,27.436340) with $B-R=2.2$ might also be the counterpart.}

\end{deluxetable}
%
%
\begin{deluxetable}{rrrrc}
\tabletypesize{\scriptsize}
\tablewidth{0pt}
\tablecolumns{5}
\tablecaption{Spectroscopically-confirmed ``Giant" Coma member galaxies \label{table1}}
\tablehead{
\colhead{Source Name} & \colhead{$z^{\rm a}$} & \colhead{L$_{X}^{\rm b}$} & \colhead{Spectral} & \colhead{NAME}  \\
\colhead{} & \colhead{}& \colhead{(10$^{39}$~\lumin)} & {ID$^{\rm c}$} \\
}
\startdata
\hline
\multicolumn{5}{c}{X-RAY-DETECTED SOURCES:} \\
\hline
  125725.25$+$272413.8 & 0.0162 & 5.3 &  e(b) & D15, GMP4918, Mkn 55 \\
  125801.54$+$272922.3 & 0.0253 & 6.6 & k    & D21, GMP4568\\
  125818.60$+$271840.0 & 0.0249 & 4.5 & e(b?)& D9,  GMP4351 \\
  125835.24$+$271552.7 & 0.0247 & 7.9 & e(a:??) & D5,  GMP4159\\
\hline
\multicolumn{5}{c}{UNDETECTED SOURCES:} \\
\hline
125736.14+272905.59 & 0.0242 & \nodata & k &  D23 \\
125754.38+272926.47 & 0.0165 & \nodata &  k+a:(e:)   \\
125833.14+272151.54 & 0.0233 & \nodata &  k &   \\

\enddata 
\tablenotetext{a}{From Mobasher et al. (2001)}
\tablenotetext{b}{These are 0.5--8~keV luminosities calculated assuming 
$H_{0}=70$~km s$^{-1}$ Mpc$^{-1}$.}
\tablenotetext{c}{From \cite{Pogg2001spectra}}
\end{deluxetable}
%
%
%


\begin{figure}[t!]
\centerline{\includegraphics[scale=1.00,angle=0]{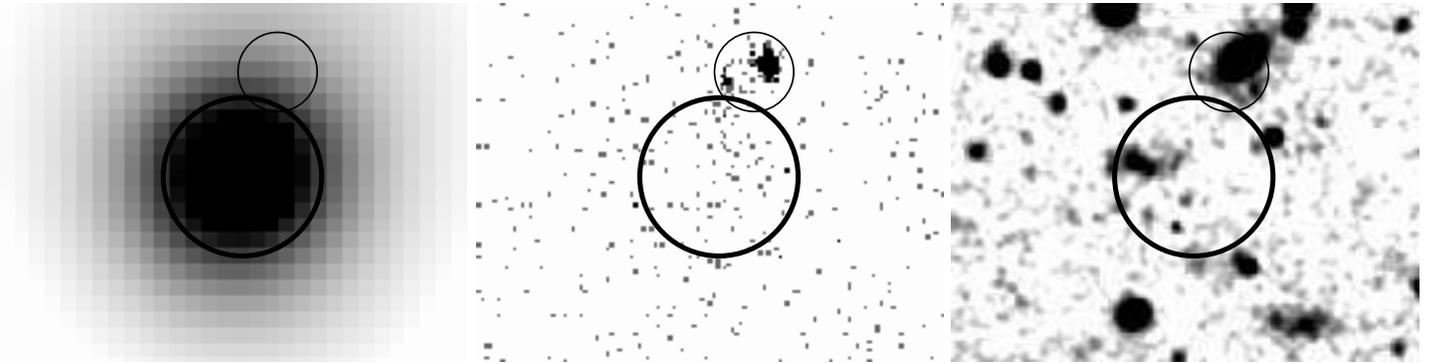}}

\caption[]{\label{diffuse_figure} Images showing the diffuse X-ray source detected in the COMA3 region (see also Figure~\ref{xray_image}).  In order, these 
are the adaptively smoothed (sources removed) 0.5--8~keV $Chandra$ image, the
raw 0.5--8~keV $Chandra$ image and the WHT $R$-band image.  The regions marked
in each image are the same, the large circle has radius 10$^{\prime \prime}$ and the small circle has radius 5$^{\prime \prime}$.  
The middle image shows that $Chandra$'s angular resolution clearly 
resolves the two nearest point sources from the diffuse region.  
This X-ray point source is clearly identified with an optically 
bright galaxy ($B=18.5$,$B-R=2.3$) for which no redshift is 
currently available.    The optically brightest source within the 10$^{\prime \prime}$ circle has $R<22$ and also has no measured spectroscopic redshift.

}

\end{figure}

\end{document}